\documentclass[
twocolumn,
]{ceurart}

\usepackage{listings}
\usepackage{tcolorbox}
\usepackage{tabularx}
\usepackage[framemethod=tikz]{mdframed}
\usepackage{multirow}
\usepackage{subcaption}
\usepackage{booktabs} 
\usepackage{array}
\usepackage{enumitem}

\definecolor{safe}{HTML}{008000}    
\definecolor{attention}{HTML}{DAA520}    
\definecolor{warning}{HTML}{FF8C00}    
\definecolor{danger}{HTML}{FF0000}    
\newcommand{\dquotes}[1]{``#1''}
\newcommand{\squotes}[1]{`#1'}

\begin{document}

\copyrightyear{2024}
\copyrightclause{Copyright for this paper by its authors.
  Use permitted under Creative Commons License Attribution 4.0
  International (CC BY 4.0).}

\conference{ROEGEN@RECSYS'24. The 1st Workshop on Risks, Opportunities, and Evaluation of Generative Models in Recommender Systems, Colocated with ACM Conference on Recommender Systems (RecSys) in Bari, Italy, October 2024}

\title{A Normative Framework for Benchmarking Consumer Fairness in Large Language Model Recommender Systems}

\tnotemark[1]


\author[]{Yashar Deldjoo}[%
orcid=0000-0002-6767-358X,
email=deldjooy@acm.org,
url=https://yasdel.github.io/,
]

\author[]{Fatemeh Nazary}[%
orcid=0000-0002-6683-9453,
email=fatemeh.nazary@poliba.it,
url=https://sisinflab.poliba.it/people/fatemeh-nazary/,
]

\address[]{Polytechnic University of Bari, Italy}

\begin{abstract}
The rapid adoption of large language models (LLMs) in recommender systems (RS) presents new challenges in understanding and evaluating their biases, which can result in unfairness or the amplification of stereotypes. Traditional fairness evaluations in RS primarily focus on collaborative filtering (CF) settings, which may not fully capture the complexities of LLMs, as these models often inherit biases from large, unregulated data. This paper proposes a normative framework to benchmark consumer fairness in LLM-powered recommender systems (RecLLMs).
We critically examine how fairness norms in classical RS fall short in addressing the challenges posed by LLMs. We argue that this gap can lead to arbitrary conclusions about fairness, and we propose a more structured, formal approach to evaluate fairness in such systems. Our experiments on the MovieLens dataset on \textit{consumer fairness,} using  in-context learning (zero-shot vs. few-shot) reveal fairness deviations in age-based recommendations, particularly when additional contextual examples are introduced (ICL-2). Statistical significance tests confirm that these deviations are not random, highlighting the need for robust evaluation methods. While this work offers a preliminary discussion on a proposed normative framework, our hope is that it could provide a formal, principled approach for auditing and mitigating bias in RecLLMs. The code and dataset used for this work will be shared at \href{your-github-link}{\textbf{gihub-anonymized}}.
\end{abstract}



\maketitle

\section{Introduction}


\noindent \textbf{Context.}  Fairness in recommender systems (RS) has garnered significant attention in recent years, driven by the need to mitigate biases that can negatively impact both consumers and providers. Most existing research in the RS field, however, has focused on fairness in classical collaborative filtering (CF) setting, which primarily rely on \textit{in-domain} user-item interaction data to compute recommendations~\cite{deldjoo2024fairness,DBLP:journals/ftir/Ekstrand0B022}. Since the introduction of ChatGPT in 2023, the RS community has seen unprecedented interest in integrating generative models—particularly those powered by pre-trained large language models (LLMs)--for personalization~\cite{deldjoo2024review,he2023large}. These models, which capture vast amounts of knowledge during pre-training on semi-supervised tasks, can be quickly adapted to a variety of contexts within RS, offering notable advantages for personalized recommendations. These benefits include \textit{efficiency} (rapid deployment and adaptability), \textit{precision and context-awareness} (enhanced personalization across diverse tasks), and \textit{robustness in data scarce scenarios} (the ability to perform well under sparse data conditions).

However, integrating LLMs into recommender systems introduces new risks, particularly biases embedded in the training data, which can lead to unfairness or the amplification of stereotypes affecting sensitive or protected groups. Given the unregulated nature of online data, it becomes a critical concern to \textit{(i)} understand, \textit{(ii)} evaluate, and \textit{(iii)} mitigate biases and unfairness of RecLLMs. RecLLMs differ significantly from traditional CF systems in several key areas: the \textit{input space} (where simple star ratings are replaced by more complex inputs like natural language user profiles), \textit{model types} (pre-trained on vast datasets rather than being directly trained using in-domain data), and \textit{output spaces} (offering more structured outputs like complementary items or detailed explanations instead of just item IDs). This work advocates for developing RecLLM fairness evaluation frameworks that account for these differences, as they can influence our understanding of what is fair and unfair. The central question explored in this work revolves around the following:

\begin{center}  
\textit{How can we audit the fairness of \textbf{RecLLMs} (recommender systems powered by large language models), and how does fairness evaluation for RecLLMs differ from traditional CF methods?}
\end{center}

\begin{figure*}
    \centering
    \includegraphics[width = 0.950\linewidth]{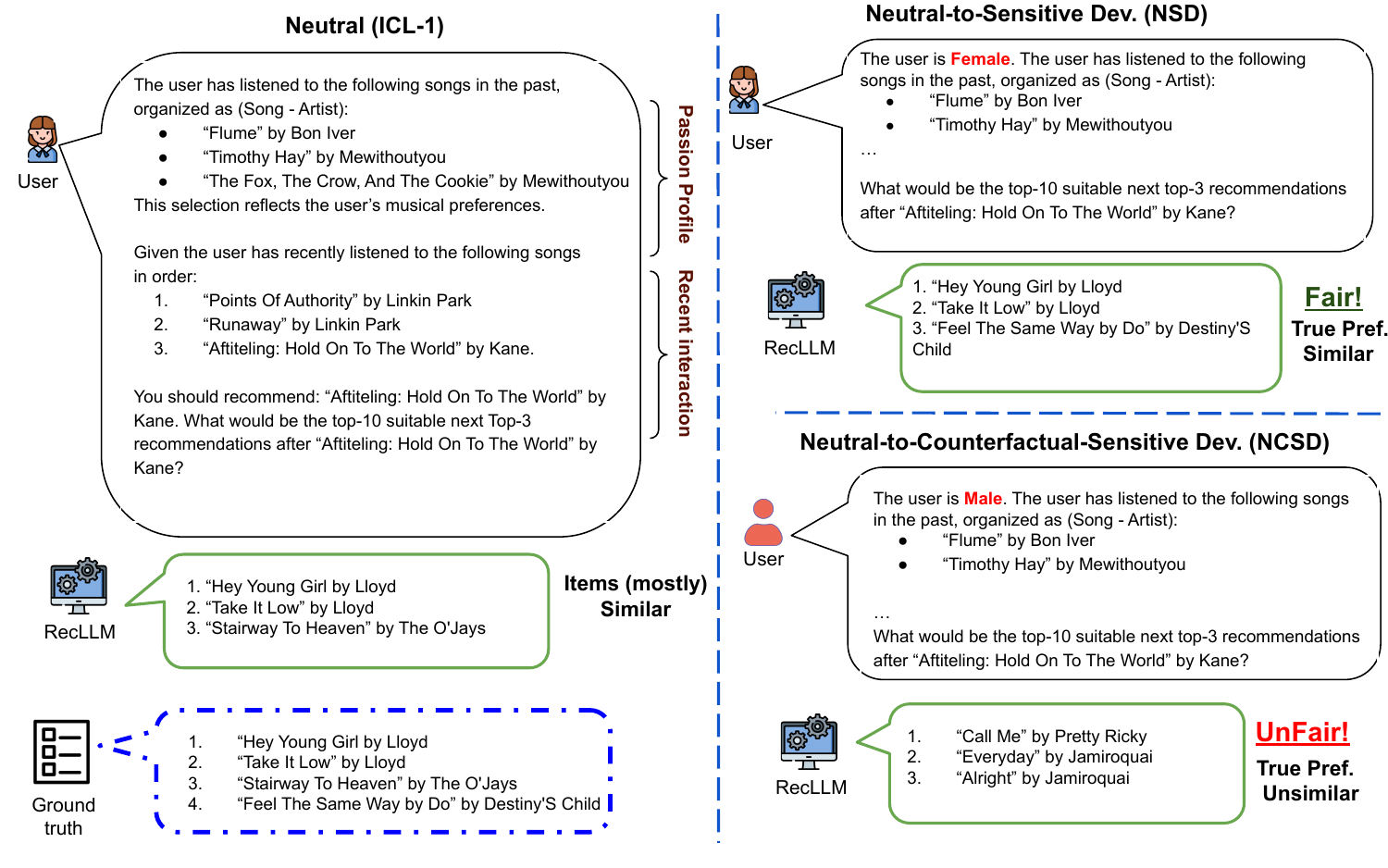}
    \caption{This figure illustrates the direct use of Large Language Models (LLMs) in generating personalized recommendations. It compares outputs under neutral conditions with those generated under scenarios that consider sensitive attributes.}
        \label{fig:fairness_evaluation}

\end{figure*}

\vspace{1mm}
\noindent \textbf{Research Problem.} In RecLLMs, there is the ability to incorporate sensitive demographic information, such as gender, directly from natural language (NL) user profiles—something that mainstream collaborative filtering (CF) models typically do not utilize. Figure~\ref{fig:fairness_evaluation} illustrates this with an example. Building on this,~\citet{zhang2023chatgpt} propose a fairness evaluation framework for RecLLMs that examines how demographic attributes, such as gender, can impact recommender outcomes. Their approach defines unfairness as the difference in recommendations between a sensitive ranker (which considers demographic factors) and a neutral ranker (which does not), based solely on differences in item IDs or their ranking in the list. This approach equates differences across user groups with unfairness, oversimplifying the issue by failing to account for situations where such differences might represent valid personalization. 

For example, if a RecLLM recommends songs like \dquotes{Hey Young Girl} by Lloyd based on prior interactions in a neutral scenario, but then suggests a song by Jamiroquai when gender is factored in, the system might flag this as unfair. This assumes the recommendation change is driven by gender stereotypes, without considering that such recommendations may align with the user's actual preferences.

To address this limitation, ~\citet{deldjoo2024cfairllm} propose the CFairLLM framework, which improves on FaiRLLM by assessing fairness based on whether the benefits a sensitive ranker provides differ (or more precisely worse) from those of a reference ranker, to flag it unfair. Thus, CFairLLM evaluates recommendation variations by comparing them to the user’s true preferences. For instance, if the sensitive ranker suggests songs that better match the user's preferences, even if they differ from the neutral ranker's list, these variations may indicate proper personalization rather than unfairness. This highlights the importance of defining fairness norms clearly; otherwise, research results might be contradictory.


Overall, these approaches lack a formal discussion of key elements essential for determining fairness in RecLLMs, such as establishing a clear reference point for evaluation and defining precise benefits and metrics for measuring fairness. 

\vspace{1mm}

\noindent \textbf{Contributions.} The current work advocates for the need for a~\dquotes{\textbf{normative framework}} to address the above issues--one that ensures fairness is evaluated according to well-established, principled standards, rather than subjective or arbitrary assumptions (e.g., assuming all differences indicate unfairness). Additionally, it highlights the importance of having clear, objective criteria for consistently and meaningfully assessing fairness.

The main contributions of this paper are as follows:

\begin{enumerate}[leftmargin=*, itemsep=1pt]
  \item \textbf{Distinguishing Fairness Frameworks:} We differentiate between two types of fairness evaluation in RecLLMs: \textit{(i)} fairness when \textit{sensitive attributes} are involved in generating recommendations (referred to as \textit{sensitive rankers}), and \textit{(ii)} fairness when comparing recommendations from neutral rankers, which do not use sensitive attributes, to predefined target distributions. This distinction is crucial for understanding the nuances in fairness evaluation between RecLLMs and traditional CF models, which typically rely on (i). Our work considers both approaches.
 
 \item \textbf{Introduction of Novel Fairness Metrics:} Building on the previous distinction, we introduce three fairness evaluation metrics, which essentially operate on the same principle: comparing the deviation and difference between the RS ranker and another ranker (either a reference or a target representation). The metrics are as follows: Neutral vs. Sensitive Ranker Deviation (NSD), Neutral vs. Counterfactual Sensitive Deviation (NCSD), and Intrinsic Fairness (IF). While NSD and NCSD are specific to RecLLMs, IF is typically applied in CF scenarios.
 
\item \textbf{Introducing of Reference Rankers:} A key aspect of fairness evaluation in RecLLMs is the use of reference rankers. Each fairness notion relies on a specific reference ranker, such as the neutral ranker (for NSD and NCSD) or a predefined target distribution (for IF). The choice of the reference ranker plays a critical role in measuring deviations, as it influences how fairness or unfairness is perceived. In contrast to previous work that focuses merely on differences between rankers, as suggested by \cite{deldjoo2024cfairllm}, we consider preference alignment with ground truth data. 

\item \textbf{Quantification of Fairness Deviations:} To quantify ranker deviation, we propose metrics that assess the quality of recommendations through both set-based and rank-based measures. These measures examine the accuracy of the ranking and the overall benefit derived by different demographic groups. In addition to evaluating benefit deviation, we apply statistical significance tests to determine whether observed differences in recommendation benefits across groups are meaningful. This provides a comprehensive mechanism to quantify fairness deviations and ensures that the quality of recommendations is assessed based on rigorous, statistically sound methods. 
\end{enumerate}
    
Overall, this work advocates the need for a~\textit{normative framework} for fairness evaluation of RecLLMs, proposing a more formal approach than previous research. In this framework, fairness is evaluated against clear, well-defined standards, aiming to avoid arbitrary assumptions and provide a structured method for assessing how sensitive information affect recommendation outcomes.
Our focus in this work is specifically on \textbf{consumer fairness}.

\section{Evaluation Framework for Consumer Fairness in RecLLMs}

We present a multi-faceted framework designed to evaluate fairness in Recommender Systems powered by Large Language Models (RecLLMs). This work builds significantly upon and extends previous research~\cite{zhang2023chatgpt, deldjoo2024cfairllm}, but offers a more formal approach to fairness evaluation in RecLLMs. 

\subsection{Definitions.}
\label{subsec:CS}


\noindent \textbf{Definition of Groups.}  In this study, fairness discussions are conducted at the group level. We denote by \( \mathcal{A} \) the set of all sensitive attributes, represented as \( \mathcal{A} = \{a, b\} \), where each attribute \( a \) and \( b \) corresponds to specific characteristics. We specifically use
\begin{itemize}
    \item The attribute \( a \) to represent \dquotes{\textbf{gender},} with the values \( a_1 \) and \( a_2 \), where \( a_1 = \text{Male} \) and \( a_2 = \text{Female} \).
    \item The attribute \( b \) to represent \dquotes{\textbf{age-groups},} with the values \( b_1 \) and \( b_2 \), where \( b_1 = \text{Young} \) and \( b_2 = \text{Old} \).
\end{itemize}
For simplicity, in this work we only consider groups that are \textbf{independent} and \textbf{binary}. However, overlapping groups can also be considered by defining combinations of attributes, such that the set of possible overlapping groups is given by \( \mathcal{G} = \{ (a_1, b_1), (a_1, b_2), (a_2, b_1), (a_2, b_2) \} \). Here, each pair represents a distinct demographic group (e.g., \( (a_1, b_1) \) for young males, \( (a_2, b_2) \) for older females), allowing to analyze fairness across intersectional identities~\cite{deldjoo2024cfairllm}.


\vspace{1mm}

\noindent \textbf{Ranking Lists.} We first introduce primary ranker types for fairness definitions.

\begin{description}

\item \textbf{Neutral Ranker} (\( \mathcal{R}_{N} \)): Referred to as the \textit{neutral ranking list}, this term describes a sequence of items \( \{i_1, i_2, \ldots, i_k\} \) ranked by a Recommender Language Learning Model (RecLLM), using NL profile (as prompts) that do not incorporate sensitive user attributes. The neutral ranker is designed to reflect scenarios based purely on non-sensitive demographic data. It bases recommendations solely on the historical interaction of the user with the system. \vspace{1.0mm}

\item \textbf{Sensitive Ranker} (\( \mathcal{R}_{S}^{a} \)): Short for \textit{sensitive ranking list}, it denotes a sequence of items \( \{i_1^a, i_2^a, \ldots, i_k^a\} \) ranked by a RecLLM using prompts that \underline{do} utilize sensitive attributes such as gender, age, etc. They aim to capture scenarios where the LLM is potentially influenced by sensitive attributes, whether positively (providing more relevant recommendations) or negatively (recommending less relevant items).

\item \textbf{Counterfactual Sensitive Ranker} (\( \mathcal{R}_{CS}^{do(a)} \)): This ranker represents a sequence of items ranked by a RecLLM under the counterfactual scenario where the sensitive attribute \( a \) is set to a specific hypothetical value through the \(do()\) operation. For example, \( \mathcal{R}_{CS}^{do(\text{Male})} \) tests the recommendations as if the gender of every user were male, regardless of their actual gender. This method allows us to explore \dquotes{\textbf{what-if}} scenarios, examining how different assumed values of sensitive attributes impact the recommendations, thereby exploring \textbf{counterfactual outcomes}. See also Section~\ref{subsec:fair_quant}, the discussion of NCSD.\footnote{Note that we recognize this might be a naive way of implementing the \dquotes{\textit{what-if}} scenario, since e.g., with \(do(\text{Male})\) and \(do(\text{Female})\), only part of the population is hypothetically altered. It nonetheless provides a framework for exploring how altering a single attribute could influence outcomes.}

\end{description}

\noindent \textbf{Fairness Frameworks.} 
\label{subsec:fair_notion}
On the consumer side, we consider the following fairness notions, each linked to the corresponding rankers:

\begin{description}
    \item[Neutral vs. Sensitive Ranker Deviation (NSD):] This notion measures disparities between the neutral ranker (\( \mathcal{R}_{N} \)) and the sensitive ranker (\( \mathcal{R}_{S}^{a} \)), evaluating how the inclusion of sensitive attributes influences the recommendations. \textit{Thus, the neutral ranker \( \mathcal{R}_{N} \) serves as the \squotes{reference} against which fairness is measured. \vspace{0.5mm}}

\item[Neutral vs. Counterfactual Sensitive Deviation (NCSD):] This concept assesses changes in recommendations when a sensitive attribute is counterfactually altered using the \(do()\) operation, setting the attribute to a specific hypothetical value. The comparison is made between the counterfactual sensitive ranker (\(\mathcal{R}_{CS}^{do(a)}\)) and the neutral ranker (\(\mathcal{R}_{N}\)). \textit{Here, we select \(\mathcal{R}_{N}\) as the reference ranker to evaluate how assumptions about changes in \(a\) affect the recommendations.}\footnote{Note that for NCSD, \(\mathcal{R}_{S}^{a}\) could also be used as the reference ranker.} \vspace{0.5mm}

    \item[Intrinsic Fairness (IF):] Focusing on qualities intrinsic to recommendations, IF evaluates the fairness of distributions generated by the neutral ranker (\( \mathcal{R}_{N} \)), and evaluates the benefits provided by the recommender across sensitive groups (e.g., male vs. female). Since no direct comparisons between sensitive ranker(s) are conducted,
    this analysis is essentially testing where the prevalence of certain sensitive groups in training data skew LLM outputs. \textit{Thus, a predefined \squotes{target distribution}, e.g., uniform, serves as the reference against which fairness is measured.} \vspace{1mm}

    \end{description}


It could be noted that both NSD and NCSD evaluate fairness across \textit{two types of rankers}, examining the potential biases introduced by sensitive attributes and their counterfactual adjustments, while IF focuses on a \textit{single ranker}, the Neutral Ranker. \vspace{2mm}

\noindent \textbf{Fairness quantification.}\label{subsec:fair_quant}
To quantify unfairness in RecLLMs, we start by defining the general concept of benefit deviation, which serves as the foundation to \textit{quantify} unfairness in our framework, given by:

\begin{equation}
    \Delta \mathcal{B} =  \mathcal{B}(\mathcal{R}_{X}) - \mathcal{B}(\mathcal{R}_{ref})  
\end{equation}
where \( \mathcal{R}_{X} \) represent ranking generated by the target recommender (e.g., sensitive ranker), \(\mathcal{R}_{ref}\) is the reference ranker (e.g., \( \mathcal{R}_{N} \) in NSD), and \( \mathcal{B} \) represents the benefit derived from each list. A lower value of \(  \Delta \mathcal{B} \) in every scenario below indicates a higher amount of unfairness. 

\begin{enumerate}
    \item \textbf{Quantifying \( \Delta \mathcal{B}  \)  for NSD.}
   \begin{equation}
    \Delta \mathcal{B} =   \mathcal{B}(\mathcal{R}_{S}^{a}) - \mathcal{B}(\mathcal{R}_{N})
   \end{equation}
   This metric compares the benefits derived from comparing 
   a sensitive ranker \( \mathcal{R}_{S}^{a} \), and a neutral ranker \( \mathcal{R}_{N} \). It evaluates how the inclusion of sensitive attributes impacts the benefits of the recommendation. A positive \( \Delta \mathcal{B}\) could indicate enhanced personalization due to the introduction of sensitive attributes, while a negative deviation could suggest unfairness due to stereotypes or biases.
\end{enumerate}

\noindent For NSD, we focus on comparing the changes across different groups, specifically:
\begin{itemize}
    \item For gender. \( \Delta \mathcal{B}_{a_1} \) and \( \Delta \mathcal{B}_{a_2} \) where \( a_1 \) and \( a_2 \) correspond to Male and Female, respectively;
    \item For age categories: \( \Delta \mathcal{B}_{b_1} \) and \( \Delta \mathcal{B}_{b_2} \) where \( b_1 \) and \( b_2 \) represent the Young and Adult groups, respectively.
\end{itemize}

We could utilize a numerical threshold (\( thr \)) set at a predefined value, to gauge the magnitude of deviations, providing a quantitative measure of potential unfairness. In our work, beyond this, we adopt a more robust approach by using statistical significance tests to measure whether the means of two distributions—specifically \( \Delta \mathcal{B}_{a_1} \) and \( \Delta \mathcal{B}_{a_2} \) for gender, and \( \Delta \mathcal{B}_{b_1} \) and \( \Delta \mathcal{B}_{b_2} \) for age categories—are significantly different. We employ the t-test for independent samples to ascertain differences between these distributions (\(p <0.05\)).\footnote{Additionally, other statistical significance tests such as the Mann-Whitney U test, a non-parametric test could be used when the data does not meet the assumptions necessary for the t-test.}

\vspace{3mm}

\noindent \textbf{Example.} Suppose the benefit deviation \( \Delta \mathcal{B}_{a_1} \) (Male) is \(0.12\), and \( \Delta \mathcal{B}_{a_2} \) (Female) is \(-0.15\). The positive deviation for males suggests an enhanced personalization effect, while the negative deviation for females, and particularly a large deviation from Male, indicates potential unfairness due to biased or stereotypical recommendations favoring males over females. 

To measure NSD, we calculate the disparity in benefit deviations as \( \delta_{\text{gender}} = \Delta \mathcal{B}_{a_1} - \Delta \mathcal{B}_{a_2}\) and \( \delta_{\text{age}} = \Delta \mathcal{B}_{b_1} - \Delta \mathcal{B}_{b_2}\), using these differences as the main measures of unfairness. We intentionally use the signed version of the metric to discern the direction of unfairness. 


\vspace{0.5mm}

\noindent \textbf{Note.} The threshold set for differentiating the levels of fairness concerns are inherently subjective and may vary depending on the specific task, system, or analysis objectives. In this work, we chose a threshold value that is reasonably suitable but acknowledge that what makes an \dquotes{appropriate} value could differ widely based on context.  Moreover, we introduce Table \ref{tab:fair_pval} to contribute to a more systematic and organized approach to categorize fairness metrics, employing \dquotes{\textbf{color coding}} to visually distinguish between the various levels of concern. 
\begin{table}[h]
\caption{Fairness Evaluation Based on the threshold $\delta$,  \( \Delta \mathcal{B} < \delta \) and p-value}
\label{tab:fair_pval}
\centering
\begin{tabular}{|c|c|c|}
\hline
\textbf{Metric} & \textbf{(\( \delta \), $p$-value)} & \textbf{Status} \\
\hline
Level 1 & Small - (p>0.05) & \textcolor{safe}{Safe} \\
Level 2 & Fairly large - (p>0.05) & \textcolor{attention}{Attention Needed} \\
Level 3 & Large - (p>0.05) & \textcolor{warning}{Likely Issue} \\
Level 4 & Large/Small - (p<0.05) & \textcolor{danger}{Significant Issue} \\
\hline
\end{tabular}
\end{table}

Table \ref{tab:fair_pval} essentially aims to present a structured assessment of fairness, organizing different levels of disparity based on \( \Delta \mathcal{B} \) and associated p-values into categories ranging from \squotes{Safe} to \squotes{Significant Issue}. This categorization helps stakeholders quickly identify potential biases in the recommendation system and determine the urgency of needed interventions. One might choose to adjust the number of levels or the criteria for each level based on their particular needs, regulations, and the nuances of their data.

\begin{enumerate}
    \item[(2)] \textbf{Neutral vs. Counterfactual Sensitive Deviation (NCSD).}
   \begin{equation}
    \Delta \mathcal{B} =   \mathcal{B}(\mathcal{R}_{CS}^{do(a)}) - \mathcal{B}(\mathcal{R}_{N})
   \end{equation}
\end{enumerate}

NCSD essentially measures the difference in recommendation performance in a Hypothetical Scenario, asking how recommendations would perform if everyone were considered to be of the same gender (e.g., male or female).  As stated before, although we could use the correct gender of the user (the sensitive ranker), we chose to use the neutral recommender as the reference. 

To explore the impact of each attribute value in a controlled, hypothetical scenario, we symbolically use the causal \(do()\) operator:
\begin{itemize}
    \item \textbf{\(do(\text{Gender} = \text{Male})\)} — Simulating the scenario where every individual, regardless of their original gender, is considered as male.
    \item \textbf{\(do(\text{Gender} = \text{Female})\)} — Simulating the scenario in which every individual is considered as female.
\end{itemize}
This method allows us to assess the outcomes if the gender of every individual was hypothetically set to Male and then to Female, (and similar for age-categories), exploring the robustness and fairness of the system under these gender-altered conditions. \vspace{0.15mm}

Finally, Intrinsic Fairness (IF) examines the fairness of recommendation distributions by a neutral ranker, \( \mathcal{R}_{N} \), \textit{across sensitive groups} such as male versus female. While the previous approaches may be more specific to RecLLMs due to the integration of \dquotes{demographic information,} IF represents a more general approach that can also be and has been widely applied to traditional recommendation models, such as collaborative filtering models \cite{boratto2022consumer,ekstrand2012fairness,li2021user}. Essentially, IF evaluates whether the outcomes provided by \( \mathcal{R}_{N} \) are fair by comparing the actual distribution of recommendations to a target (uniform) distribution across different demographic groups.

\subsubsection{Benefit types.} To provide a nuanced assessment of the benefits derived from recommendations, we implement two specific measures:

\begin{description}
\item \textbf{Hit (\( \mathcal{B}_{hit} \)).} Measures whether the items in a recommendation list are relevant to the user. Specifically, the hit rate evaluates if any of the top \( k \) items recommended by the system appear in the ground truth list of user preferred items.

\item \textbf{Ranking Quality (\( \mathcal{B}_{rank} \)).} Assesses the alignment between the order of items in the recommendation list and their actual relevance to the user, as determined by their position in the ground truth list. This metric indicates how effectively the recommendation system orders items in a way that corresponds to the user preferences.
 \end{description}

These metrics serve as specific instances of \( \mathcal{B} \) in our framework, allowing us to measure the practical benefits of the recommendations provided by different RecLLM scenarios. 

\section{Experiment}

\subsection{Setup}

We conducted a series of experiments to evaluate the fairness and effectiveness of recommendations generated by our proposed RecLLM fairness evaluation framework. The experiments focused on two key aspects: (i) understanding the behavior of the model when no prior examples are available (0-shot learning) versus (ii) the effect of providing one (ICL-1) and two examples (ICL-2) in context for generating recommendations.

\begin{table}[h]
\centering

\caption{Summary statistics of the LastFM and MovieLens datasets after filtering. \(|R|\) represents the number of interactions, while \(|R|/|U|\) denotes the average number of interactions per user \(|U|\). The training and testing data statistics are shown for each dataset.}
\label{tab:dataset-summary}
\begin{tabular}{@{}lcc@{}}
\toprule
\textbf{Dataset / Statistic} & \textbf{Training Data} & \textbf{Testing Data} \\
\midrule
\textbf{MovieLens} & \begin{tabular}[c]{@{}c@{}}$|R| = 16,757$\\|R|/|U| = 209.46\end{tabular} & \begin{tabular}[c]{@{}c@{}}$|R| = 4,230$\\|R|/|U| = 52.88\end{tabular} \\
\bottomrule
\end{tabular}
\end{table}

The experiments utilize two datasets, LastFM and MovieLens, which offer a combination of music and movie recommendation tasks. However, due to space constraints, we report only the results for the MovieLens dataset. We follow the procedure outlined in \cite{deldjoo2024understanding,deldjoo2024cfairllm} to build a sequential recommender system focused on next-item prediction.

We evaluated the model performance based on fairness metrics related to gender and age group, using both Neutral vs. Sensitive Ranker Deviation (NSD) and Neutral vs. Counterfactual Sensitive Deviation (NCSD). These metrics help quantify how incorporating sensitive attributes, such as gender and age, affects the fairness and relevance of the recommendation system.

Our approach involves a sequential recommendation task where we employ timestamps to ensure the data is split temporally. Initially, we randomly select a subset of 80 users who exhibit a moderate level of interaction within the datasets. This allows us to handle the data efficiently while ensuring that the users selected have enough interactions to inform the training process but are not so many as to skew the representativeness of typical user behavior. The data for these users is then divided into training and test sets by sorting their interactions over time and splitting them such that 80\% of a user's interactions are used for training, with the remaining 20\% held out for testing. This method respects the chronological order of interactions, thereby simulating a realistic scenario where a model can only learn from past data to make predictions about future user behavior.

We assessed model performance across various conditions:
\begin{itemize}
    \item \textbf{0-shot learning}: No examples of past recommendations are provided to guide the system.
    \item \textbf{ICL-1}: One example of past user interaction is provided to improve contextual understanding.
    \item \textbf{ICL-2}: Two examples are provided to further enhance recommendation relevance and fairness.
\end{itemize}

The main goal is to test the extent to which including sensitive attributes (gender and age) and providing in-context examples influences both the fairness and relevance of the recommendations generated by the model.

The recommendation generation was tested under different strategies for profile sampling stragies:
\begin{itemize}
    \item \textbf{rand}: : Uniform selection of tracks or movies to provide a stochastic view of user preferences.
    \item \textbf{freq}: Prioritization of tracks/movies based on their frequency of playback, and rating provided emphasizing the main preferences of the user
    \item \textbf{rec-freq}: This hybrid approach combines the recency and frequency of track interactions using a weighted score formula. 
\end{itemize}

To save space, we present a snapshot of the results as initial support for our framework using the MovieLens dataset. A more detailed extension will be provided later. Note that the recommendation scenario here is for \textit{sequential item recommendation} task.

\subsection{Results and Discussion}

Table \ref{tab:fairness_results_final} shows the fairness evaluation results across various conditions. Key findings include:

\begin{table}[h]
\centering
\caption{Key Fairness Results for Gender and Age Groups with NSD and NCSD, Including $\Delta B_1$ and $\Delta B_2$}
\label{tab:fairness_results_final}
\small
\begin{tabular}{|l|c|c|c|}
\hline
\multicolumn{4}{|c|}{\textbf{NSD/Gender}} \\ \hline
\textbf{Condition}    & \(\Delta B_1\)   & \(\Delta B_2\)   & \(\delta_{\text{gender}} (p\text{-value})\)  \\ \hline
0-shot/rand           & -0.0074          & -0.019           & 0.0116 (p=0.730)                     \\ \hline
ICL-1/rand            & -0.0222          & -0.0476          & 0.0254 (p=0.464)                     \\ \hline
ICL-2/rand            & -0.037           & -0.019           & -0.018 (p=0.386)                     \\ \hline
0-shot/freq           & 0.0              & -0.019           & 0.019 (p=0.609)                      \\ \hline
ICL-1/freq            & 0.0148           & -0.0095          & 0.0243 (p=0.368)                     \\ \hline
ICL-2/freq            & 0.0148           & -0.019           & 0.0339 (p=0.425)                     \\ \hline
\hline
\multicolumn{4}{|c|}{\textbf{NCSD/Age-Group}} \\ \hline
\textbf{Condition}    & \(\Delta B_1\)   & \(\Delta B_2\)   & \(\delta_{\text{age-gr}} (p\text{-value})\) \\ \hline
0-shot/rand           & 0.0              & -0.0046          & 0.0046 (p=0.954)                     \\ \hline
ICL-1/rand            & 0.0              & -0.0365          & 0.0365 (p=0.812)                     \\ \hline
ICL-2/rand            & 0.0952           & -0.0228          & \textcolor{orange}{\textbf{0.1181 (p=0.108)}} \\ \hline
0-shot/freq           & 0.0476           & -0.0091          & 0.0568 (p=0.307)                     \\ \hline
ICL-1/freq            & 0.0476           & -0.0091          & 0.0568 (p=0.312)                     \\ \hline
ICL-2/freq            & 0.0              & -0.0548          & \textcolor{red}{\textbf{0.0548 (p=0.022)}}    \\ \hline
\end{tabular}
\end{table}
\begin{itemize}
    \item \textbf{Gender-Based Fairness (NSD).} Gender fairness was mostly stable across conditions, with minor deviations observed. The most noticeable case was under the \textbf{ICL-2/freq} condition, where the system slightly favored one gender group (i.e., males)  (\(\delta = 0.0339\), p = 0.425). While not statistically significant, this result suggests the model may introduce slight gender biases when more contextual examples are provided.
   \item \textbf{Age-Based Fairness (NCSD).} Age fairness showed more pronounced issues. Under the \textbf{ICL-2/freq} condition, the deviation was statistically significant (\(\delta = 0.0548\), p = 0.022), indicating the system significantly favored one age group (Old) when two contextual examples were used. Similarly, \textbf{ICL-2/rand} displayed a notable deviation (\(\delta = 0.1181\), p = 0.108), though it was not statistically significant.
\item {Impact of Contextual Information.} As more contextual examples were introduced (moving from 0-shot to ICL-1 and ICL-2), deviations became more pronounced, particularly for age groups. This indicates that while context improves recommendation relevance, it can also exacerbate biases.

\end{itemize}

In conclusion, while the system demonstrates relatively strong performance in terms of gender fairness, concerns remain regarding age-based fairness, particularly in the \textbf{ICL-2/freq} condition. The experiments related to IF focus on measuring the fairness of consumers using sensitive attributes; details on these experiments will be provided in future work. Our primary objective here is not to present detailed experiments on all the elements used in this study (such as ICL, profile sampling type, or sensitive attributes), but rather to illustrate the effectiveness of the proposed framework through empirical analysis.

\section{Conclusion}
In this paper, we introduced a normative framework for benchmarking consumer fairness in large language model (LLM)-based recommender systems (RecLLMs), addressing the limitations of traditional fairness evaluations applied to collaborative filtering models. We provide a more formal and structured approach to auditing fairness by introducing key elements such as Neutral vs. Sensitive Ranker Deviation (NSD), Neutral vs. Counterfactual Sensitive Deviation (NCSD), and Intrinsic Fairness (IF). These metrics offer a principled way to assess fairness by clearly defining the reference point for fairness evaluation, whether it is a neutral ranker or a target distribution, and by quantifying fairness deviations through statistical tests. Additionally, we highlight the importance of specifying the underlying benefit types, such as hit rate and ranking quality, which provide a clear foundation for measuring fairness in relation to user preferences.

Our experiments on the MovieLens dataset demonstrate that while fairness remains stable in gender-based groups, age-based fairness deviations become more pronounced, especially when contextual examples are introduced (ICL-2). This suggests a potential amplification of biases when more contextual information is provided to the model. Future work should focus on refining these formal metrics, expanding the framework to cover more diverse sensitive attributes, and exploring further strategies to mitigate bias

\bibliography{refs}

\begin{thebibliography}{10}
\expandafter\ifx\csname natexlab\endcsname\relax\def\natexlab#1{#1}\fi
\providecommand{\url}[1]{\texttt{#1}}
\providecommand{\href}[2]{#2}
\providecommand{\path}[1]{#1}
\providecommand{\DOIprefix}{doi:}
\providecommand{\ArXivprefix}{arXiv:}
\providecommand{\URLprefix}{URL: }
\providecommand{\Pubmedprefix}{pmid:}
\providecommand{\doi}[1]{\href{http://dx.doi.org/#1}{\path{#1}}}
\providecommand{\Pubmed}[1]{\href{pmid:#1}{\path{#1}}}
\providecommand{\bibinfo}[2]{#2}
\ifx\xfnm\relax \def\xfnm[#1]{\unskip,\space#1}\fi
\bibitem[{Deldjoo et~al.(2024)Deldjoo, Jannach, Bellogin, Difonzo, and Zanzonelli}]{deldjoo2024fairness}
\bibinfo{author}{Y.~Deldjoo}, \bibinfo{author}{D.~Jannach}, \bibinfo{author}{A.~Bellogin}, \bibinfo{author}{A.~Difonzo}, \bibinfo{author}{D.~Zanzonelli},
\newblock \bibinfo{title}{Fairness in recommender systems: research landscape and future directions},
\newblock \bibinfo{journal}{User Modeling and User-Adapted Interaction} \bibinfo{volume}{34} (\bibinfo{year}{2024}) \bibinfo{pages}{59--108}.
\bibitem[{Ekstrand et~al.(2022)Ekstrand, Das, Burke, and Diaz}]{DBLP:journals/ftir/Ekstrand0B022}
\bibinfo{author}{M.~D. Ekstrand}, \bibinfo{author}{A.~Das}, \bibinfo{author}{R.~Burke}, \bibinfo{author}{F.~Diaz},
\newblock \bibinfo{title}{Fairness in information access systems},
\newblock \bibinfo{journal}{Found. Trends Inf. Retr.} \bibinfo{volume}{16} (\bibinfo{year}{2022}) \bibinfo{pages}{1--177}. \URLprefix \url{https://doi.org/10.1561/1500000079}. \DOIprefix\doi{10.1561/1500000079}.
\bibitem[{Deldjoo et~al.(2024)Deldjoo, He, McAuley, Korikov, Sanner, Ramisa, Vidal, Sathiamoorthy, Kasirzadeh, and Milano}]{deldjoo2024review}
\bibinfo{author}{Y.~Deldjoo}, \bibinfo{author}{Z.~He}, \bibinfo{author}{J.~McAuley}, \bibinfo{author}{A.~Korikov}, \bibinfo{author}{S.~Sanner}, \bibinfo{author}{A.~Ramisa}, \bibinfo{author}{R.~Vidal}, \bibinfo{author}{M.~Sathiamoorthy}, \bibinfo{author}{A.~Kasirzadeh}, \bibinfo{author}{S.~Milano},
\newblock \bibinfo{title}{A review of modern recommender systems using generative models (gen-recsys)},
\newblock in: \bibinfo{booktitle}{Proceedings of the 30th ACM SIGKDD Conference on Knowledge Discovery and Data Mining}, \bibinfo{year}{2024}, pp. \bibinfo{pages}{6448--6458}.
\bibitem[{He et~al.(2023)He, Xie, Jha, Steck, Liang, Feng, Majumder, Kallus, and McAuley}]{he2023large}
\bibinfo{author}{Z.~He}, \bibinfo{author}{Z.~Xie}, \bibinfo{author}{R.~Jha}, \bibinfo{author}{H.~Steck}, \bibinfo{author}{D.~Liang}, \bibinfo{author}{Y.~Feng}, \bibinfo{author}{B.~P. Majumder}, \bibinfo{author}{N.~Kallus}, \bibinfo{author}{J.~McAuley},
\newblock \bibinfo{title}{Large language models as zero-shot conversational recommenders},
\newblock in: \bibinfo{booktitle}{Proceedings of the 32nd ACM international conference on information and knowledge management}, \bibinfo{year}{2023}, pp. \bibinfo{pages}{720--730}.
\bibitem[{Zhang et~al.(2023)Zhang, Bao, Zhang, Wang, Feng, and He}]{zhang2023chatgpt}
\bibinfo{author}{J.~Zhang}, \bibinfo{author}{K.~Bao}, \bibinfo{author}{Y.~Zhang}, \bibinfo{author}{W.~Wang}, \bibinfo{author}{F.~Feng}, \bibinfo{author}{X.~He},
\newblock \bibinfo{title}{Is chatgpt fair for recommendation? evaluating fairness in large language model recommendation},
\newblock in: \bibinfo{booktitle}{Proceedings of the 17th ACM Conference on Recommender Systems}, \bibinfo{year}{2023}, pp. \bibinfo{pages}{993--999}.
\bibitem[{Deldjoo and Di~Noia(2024)}]{deldjoo2024cfairllm}
\bibinfo{author}{Y.~Deldjoo}, \bibinfo{author}{T.~Di~Noia},
\newblock \bibinfo{title}{Cfairllm: Consumer fairness evaluation in large-language model recommender system},
\newblock \bibinfo{journal}{arXiv preprint arXiv:2403.05668}  (\bibinfo{year}{2024}).
\bibitem[{Boratto et~al.(2022)Boratto, Fenu, Marras, and Medda}]{boratto2022consumer}
\bibinfo{author}{L.~Boratto}, \bibinfo{author}{G.~Fenu}, \bibinfo{author}{M.~Marras}, \bibinfo{author}{G.~Medda},
\newblock \bibinfo{title}{Consumer fairness in recommender systems: Contextualizing definitions and mitigations},
\newblock in: \bibinfo{booktitle}{European Conference on Information Retrieval}, \bibinfo{organization}{Springer}, \bibinfo{year}{2022}, pp. \bibinfo{pages}{552--566}.
\bibitem[{Ekstrand et~al.(2012)Ekstrand, Das, Burke, and Diaz}]{ekstrand2012fairness}
\bibinfo{author}{M.~D. Ekstrand}, \bibinfo{author}{A.~Das}, \bibinfo{author}{R.~Burke}, \bibinfo{author}{F.~Diaz},
\newblock \bibinfo{title}{Fairness in recommender systems},
\newblock in: \bibinfo{booktitle}{Recommender systems handbook}, \bibinfo{publisher}{Springer}, \bibinfo{year}{2012}, pp. \bibinfo{pages}{679--707}.
\bibitem[{Li et~al.(2021)Li, Chen, Fu, Ge, and Zhang}]{li2021user}
\bibinfo{author}{Y.~Li}, \bibinfo{author}{H.~Chen}, \bibinfo{author}{Z.~Fu}, \bibinfo{author}{Y.~Ge}, \bibinfo{author}{Y.~Zhang},
\newblock \bibinfo{title}{User-oriented fairness in recommendation},
\newblock in: \bibinfo{booktitle}{Proceedings of the web conference 2021}, \bibinfo{year}{2021}, pp. \bibinfo{pages}{624--632}.
\bibitem[{Deldjoo(2024)}]{deldjoo2024understanding}
\bibinfo{author}{Y.~Deldjoo},
\newblock \bibinfo{title}{Understanding biases in chatgpt-based recommender systems: Provider fairness, temporal stability, and recency},
\newblock \bibinfo{journal}{ACM Transactions on Recommender Systems}  (\bibinfo{year}{2024}).

\end{thebibliography}

\end{document}